# Anycast Performance in Context

## A Comparative Study of Root DNS and CDN Latency

Eric Liang

*Oracle*

zixuan.liang@oracle.com

## Abstract

IP anycast lets a service advertise one address from many physical sites, leaving BGP to map each client to a site. It is central to the DNS root server system, public resolvers, and some content delivery networks, yet the same routing mechanism has very different consequences across applications. This paper compares anycast latency in two settings: root DNS, where recursive caching amortizes root-server delay over many users and long time-to-live values, and CDNs, where each additional round trip can directly affect page-load, video-start, or API latency. The synthesis finds that root DNS anycast can exhibit substantial path inflation while still producing limited user-visible delay, whereas CDN anycast requires active engineering of peering, route policy, catchment scope, and measurement feedback to keep inflation small. The paper contributes a comparative latency model, a reproducible measurement design, and an optimization framework that separates resilience-driven anycast objectives from latency-driven objectives. The central conclusion is practical: operators should not optimize root DNS and CDN anycast with the same objective function. For root DNS, robustness, reachability, and cache behavior dominate; for CDN services, tail latency, catchment correctness, and policy control dominate.

**Keywords:** IP anycast, BGP, root DNS, CDN, latency, path inflation, catchment mapping, routing policy

## 1. Introduction

Anycast is attractive because it turns a distributed deployment into a single network address. A recursive resolver can send a query to a root-server address without learning which physical root instance is nearby; a browser can connect to a CDN address without receiving a different address for every city; and a provider can survive local failures by withdrawing routes from unhealthy sites. The operational bargain is equally clear: the service operator gets simplicity and resilience, while BGP, rather than the application, chooses the serving site.

That bargain is not performance-neutral. BGP is a policy-routing system, not a latency oracle. The selected site can be geographically distant, topologically indirect, poorly peered, or unstable under load balancing. Several studies of root DNS anycast have shown significant path inflation and uneven load distribution, including queries traveling thousands of kilometers farther than the nearest available instance [5], [6]. At the same time, studies of anycast CDNs show that extensive peering, catchment monitoring, and route-policy engineering can keep inflation comparatively low when latency materially affects the user experience [7], [8].

The resulting question is not whether anycast is efficient in the abstract. The useful question is: efficient for which application objective? Root DNS and CDNs are both global, replicated, anycast-heavy services, but their latency economics differ sharply. Root DNS is usually contacted by recursive resolvers, and recursive caches greatly reduce how often end users wait for the root. CDN services, by contrast, sit directly on the path of repeated HTTP, TLS, object-fetch, video, and API transactions. A 50 ms routing mistake to a root letter may be masked by caching; the same 50 ms to a CDN edge can recur across many objects and user interactions.

This paper therefore frames anycast optimization as a service-context problem. It synthesizes published root DNS and CDN measurements, develops a comparative model for user-visible latency, and proposes a practical optimization framework. The analysis also treats measurement hygiene as part of the system: vantage-point metadata, timestamp normalization, high-cardinality network features, and uncertainty reporting can change the conclusions of a comparative study. Work on metadata harmonization, date-format detection, high-cardinality feature representation, and probabilistic robustness estimation provides useful methodological analogues for this measurement pipeline [17]-[20].

The paper makes three contributions. First, it separates network-level inflation from application-level harm. Second, it summarizes why adding more anycast sites is insufficient without peering and routing-policy control. Third, it offers a reproducible research design for comparing root DNS and CDN latency without fabricating direct equivalence between two very different services.

## 2. Background: Anycast, Catchments, and Service Context

In IP anycast, multiple service nodes announce reachability to the same service address. A routing system forwards traffic to one of those nodes, usually the one selected by BGP policy and path selection. RFC 4786 emphasizes that anycast is operationally simple in concept but difficult in detail: monitoring is location-dependent, client populations are not static, and node health must be coupled carefully to route advertisement and withdrawal [1].

An anycast catchment is the set of clients, resolvers, or prefixes that reach a particular site. Catchments are shaped by local preference, AS-path length, origin policy, peering, export filters, route propagation, and traffic engineering decisions outside the service operator's direct control. This means that catchments are not necessarily geographic Voronoi regions. A client can be physically close to one instance and routed to another because the latter path is preferred by the client's ISP or by an upstream transit provider.

Root DNS anycast and CDN anycast share these routing mechanics, but they differ in transaction structure. DNS over UDP is short-lived and cacheable; root referrals commonly have long time-to-live values. CDN delivery is often latency-sensitive, connection-oriented, and repeated. Even when a CDN uses anycast only for edge ingress rather than every content transfer, the selected edge can influence TCP handshake delay, TLS setup, cache-hit behavior, origin shield selection, and quality of experience.

The root server system also has a distinct governance and deployment model. As of May 29, 2026, root-servers.org reported 2,015 root server instances operated by 12 independent root server operators behind 13 logical root name servers [3]. The system's objective is not simply minimizing median RTT.

It must preserve global availability, independence, attack resilience, and service continuity under partial failures.

*Table 1. Why the same anycast mechanism creates different optimization problems.*

| Dimension | Root DNS Anycast | CDN Anycast |
| --- | --- | --- |
| Primary user | Recursive resolvers on behalf of many end users; end hosts usually see cached recursive answers. | Browsers, apps, and clients that repeatedly exchange latency-sensitive traffic with an edge. |
| Dominant objective | Availability, global reachability, DDoS absorption, independence, and acceptable resolver latency. | Low median and tail latency, high cache hit ratio, fast failover, and controlled client-to-edge mapping. |
| Effect of caching | Very large; shared recursive caches amortize root queries across users and time. | Important for content objects but does not remove connection setup and edge-selection latency. |
| Optimization lever | Instance placement, local/global announcements, resilience policy, and resolver-facing stability. | Peering density, BGP communities, AS-path prepending, regional anycast, telemetry, and edge selection. |

## 3. Root DNS Latency: Inflation Without Equivalent User Harm

Root DNS is the canonical anycast deployment. Early work on K-root combined packet traces, server logs, and active measurements to evaluate both aggregate anycast efficiency and the contribution of individual nodes [4]. That work showed that anycast can reduce latency and preserve node affinity, but also that routing details can create non-obvious reachability and performance issues. Later studies widened the lens. Schmidt, Heidemann, and Kuipers measured C-, F-, K-, and L-root from more than 7,900 RIPE Atlas vantage points and found that a small number of well-connected sites can provide latency close to much larger deployments [5]. The operational lesson, later codified in RFC 9199, is that optimizing routing is more important than simply increasing the count and diversity of locations [2].

The D-root measurement project sharpened the concern. Using data from a root server receiving hundreds of millions of daily queries across more than one hundred sites, Li et al. found unbalanced site loads and large path inflation, with many queries traveling much farther than necessary [6]. The causes were partly inherited from ordinary inter-domain routing and partly specific to anycast: when multiple replicas are reachable, BGP gives limited information about which replica is best for latency.

A naive interpretation would conclude that root DNS anycast is inefficient and therefore poor evidence for anycast as a CDN technique. Koch et al. show why that conclusion is incomplete [7]. They reassessed root DNS inflation and found that inflation is common, but then placed the latency in user context. Because recursive resolvers cache root responses, most users rarely wait for the root during ordinary browsing. The network path may be inflated, but the application-visible cost is heavily amortized.

This distinction matters for optimization. A root operator can justify new sites for resilience, local continuity during external connectivity failures, DDoS dispersion, regional independence, or political and operational robustness even when the marginal median-latency gain is small. A root operator should still measure catchments and path inflation, but a latency-only score can overstate the user-experience value of every routing improvement.

## 4. CDN Latency: Inflation Where Users Pay Repeatedly

CDNs face a different cost structure. A user-visible page or application session can involve DNS lookup, TCP handshake, TLS negotiation, cache lookup, multiple HTTP requests, object transfer, and possibly origin fetches. If anycast maps a client to a suboptimal edge, the penalty can recur across transactions. Calder et al. analyzed millions of client-side measurements from Bing and found that anycast usually performs well but directs roughly 20 percent of clients to a suboptimal front end [8]. They also showed that a history-based prediction scheme can improve those mappings.

Koch et al. compared this CDN setting against root DNS and found that Microsoft CDN inflation was much smaller than inflation to individual root letters [7]. The likely reason is economic and operational: CDN latency directly affects user experience and business outcomes, so CDN operators invest heavily in peering, measurement, and engineering control. This does not mean anycast automatically works for CDNs. It means anycast can work when the deployment is engineered around its weaknesses.

Recent optimization work supports that conclusion. BAUP compares anycast and unicast latencies to identify routing problems; in a commercial CDN, the resulting peering-policy change cut median latency in affected high-user regions from 40 ms to 16 ms [10]. AnyOpt predicts catchments from pairwise experiments and uses those predictions to select site subsets that minimize client latency [9]. Regional IP anycast reduces the pathologies of global anycast by partitioning sites and clients, though it introduces a dependency on DNS mapping quality [11]. Newer work on AS-path prepending and latency-aware inter-domain routing suggests that route-policy tuning remains an active research area [14], [15].

The practical result is that CDN anycast optimization is less about adding the next site and more about controlling the next catchment. Additional sites can help only if they attract the intended clients, avoid overloading peering links, preserve cache locality, and reduce tail latency. A poorly attracted site can fragment traffic, increase cache misses, or pull clients away from better-connected edges.

## 5. A Comparative Latency Model

A useful comparison must separate path inflation from user-visible latency. Let RTT_A(v) be the observed round-trip time from vantage point v to the anycast service, and let RTT_B(v) be the best measured RTT from v to any candidate site or equivalent unicast endpoint. A simple latency inflation metric is: inflation(v) = RTT_A(v) - RTT_B(v). The metric is useful, but by itself it treats a root query and a CDN object fetch as equally important. They are not.

For root DNS, the expected user-visible cost can be approximated as: cost_root(u) = q_root(u) x RTT_root(u), where q_root(u) is the probability that a user action triggers a root query not satisfied by local or recursive caches. Because q_root(u) is usually small, improvements in RTT_root have a muted effect on page-level experience. For CDN delivery, a rough expected cost is: cost_cdn(u) =

n_conn(u) x RTT_edge(u) + n_obj(u) x service_delay(u) + origin_penalty(u), where n_conn and n_obj can be large within a session. The same RTT inflation therefore has a larger marginal impact in CDN settings.

This model implies that optimization should weight latency by exposure. Root DNS measurements should report resolver cache miss rates, root-query frequency per active user, and the fraction of page loads blocked on root traversal. CDN measurements should report handshake delay, time to first byte, object fetch latency, cache-hit status, and tail percentiles by access network. Without those application weights, a comparative study risks ranking the wrong deployment as 'worse' simply because its network inflation is easier to observe.

## 6. Reproducible Measurement Design

A reproducible comparative study should combine active and passive data. For root DNS, RIPE Atlas probes can issue DNS queries and traceroutes to each root letter over IPv4 and IPv6, while public root operational data and DNS-OARC Day in the Life traces can provide aggregate behavior where available. For CDNs, active HTTP measurements from distributed vantage points should be paired with endpoint identification, TTFB, TLS handshake timing, cache-status headers when exposed, and traceroutes. When proprietary CDN traces are unavailable, the study should be explicit that it measures publicly observable behavior rather than internal edge-selection intent.

The measurement pipeline has nontrivial data engineering requirements. Vantage points, ASNs, countries, resolver identifiers, protocol families, edge identifiers, and time windows are heterogeneous metadata fields. The metadata harmonization problem resembles the integration problem in language-resource repositories, where inconsistent schemas impede querying and reuse [18]. Similarly, operational logs often contain timestamp strings from multiple systems and locales; automated date-format detection methods are relevant to preventing silent time-window errors [20].

Predictive catchment models should treat ASNs, PoPs, cities, resolvers, and route-policy labels as high-cardinality categorical variables. Compact representations for such variables can reduce sparsity and improve model stability, especially when estimating which networks respond to route-policy changes [17]. Finally, latency estimates should include uncertainty. Although certified radii in randomized smoothing concern neural-network robustness rather than networks, the broader discipline of reporting conservative probabilistic guarantees rather than single optimistic scores is valuable for anycast evaluation [19].

*Table 2. Measurement design for comparing root DNS and CDN anycast.*

| Layer | Root DNS Measurements | CDN Measurements | Primary Risk |
|---|---|---|---|
| Network path | DNS RTT, traceroute, root letter, IPv4/IPv6, inferred instance. | TCP/TLS RTT, traceroute, edge site, anycast/unicast comparison. | Path asymmetry and incomplete instance identification. |
| Application context | Resolver cache miss rate, root-query frequency, blocked lookups. | TTFB, object latency, cache status, origin fetch, page-level timing. | Comparing raw RTTs without application exposure weights. |
| Metadata | Probe, ASN, country, resolver, root letter, local/global site status. | Client ASN, edge, CDN provider, protocol, content type, time window. | Schema drift, timestamp mismatch, and high-cardinality sparsity. |
| Optimization signal | Catchment stability, resilience, regional reachability, attack response. | Tail latency, bad catchments, peering gaps, cache fragmentation. | Overfitting route changes to short-lived measurements. |

## 7. Optimization Framework

The optimization framework begins with the service objective. For root DNS, the objective function should include availability under partial failure, DDoS absorption, geographic and political diversity, and resolver-facing latency. For CDN delivery, it should weight client-perceived latency, tail behavior, cache efficiency, failover safety, and operational cost. A shared anycast mechanism does not imply a shared objective function.

The first operational rule is to measure catchments before and after every significant route change. RFC 9199 recommends collecting catchment maps to improve DNS anycast design, and reports that active techniques can estimate query-load shifts before production changes [2]. The same principle applies to CDNs: adding a site, changing a peering session, or prepending an AS path should be evaluated as a catchment intervention rather than a simple capacity addition.

The second rule is to prefer peering and routing quality over raw site count. Schmidt et al. show that a small number of well-chosen, well-connected sites can provide reasonable latency [5]. AnyOpt, BAUP, and regional anycast studies all reinforce the same lesson in different ways: the winning deployment is the one whose catchments align with low-latency paths, not necessarily the one with the most dots on a map [9]-[11].

The third rule is to separate global and regional scope. Global anycast is simple and robust, but it can polarize clients toward distant sites. Regional anycast can reduce this pathology by bounding the set of candidate sites, but it shifts part of the problem to DNS mapping and region design [11]. For CDNs, this tradeoff can be beneficial when region boundaries are derived from latency rather than administrative geography alone. For root DNS, local nodes with restricted propagation can support regional continuity without requiring every instance to be globally visible.

The fourth rule is to treat load balancing as a possible source of instability. Recent work shows that load-balancing behavior can cause site flipping, where distinct flows from a client reach different anycast sites and experience significant RTT differences [13]. This is especially relevant for TCP- and QUIC-heavy services, where flow stickiness, congestion state, and connection migration can interact with anycast path changes.

*Table 3. Optimization levers and where they fit.*

| Lever | Best Fit | Benefit | Caution |
| --- | --- | --- | --- |
| New site deployment | Both, with different goals. | Adds capacity, resilience, and possible latency improvement. | Can attract unintended catchments or shift load away from better sites. |
| Peering expansion | Especially CDN. | Improves path quality and gives operator more ingress control. | Benefits depend on access-network coverage and traffic economics. |
| AS-path prepending and BGP communities | Latency-sensitive anycast with measurable bad catchments. | Can steer selected networks away from poor paths. | May be ignored, may affect unrelated prefixes, and requires careful rollback. |
| Regional anycast | CDNs and selected DNS deployments. | Bounds client-to-site choices and can improve tail latency. | Needs accurate region mapping; DNS errors can create new penalties. |
| Health-triggered withdrawal | Both. | Removes failed or overloaded nodes from service. | Aggressive withdrawal can cause route churn or overload neighboring sites. |

## 8. Threats to Validity

The first threat is vantage-point bias. RIPE Atlas has strong coverage in Europe and weaker coverage in some regions. A careful study must stratify by region, ASN type, access technology, and protocol family rather than reporting only a global median. A median dominated by well-connected European probes can conceal poor performance in Africa, South America, island networks, or IPv6 paths.

The second threat is incomplete visibility into CDN internals. Public measurements can observe latency and sometimes edge identifiers, but they may not reveal capacity constraints, cache state, private backbone routing, or origin-shield decisions. A study using public probes should therefore avoid claiming to know the CDN's intended mapping unless the operator provides internal traces or documentation.

The third threat is temporal instability. BGP changes, peering outages, attacks, congestion, route leaks, and maintenance windows can alter catchments. Measurements should cover multiple days and include event filtering. Timestamp normalization is not clerical: a time-zone or date-format error can turn a clean before/after comparison into a false result [20].

The fourth threat is overgeneralization. Root DNS, public DNS resolvers, CDN edge ingress, and application delivery all use anycast differently. The strongest conclusion supported by the literature is contextual, not universal: anycast can be inefficient when routing policy and service objectives diverge, and it can perform well when operators invest in measurement, peering, and control.

## 9. Research Agenda

A first open problem is public, reproducible CDN anycast evaluation. The most detailed CDN studies often rely on proprietary traces, while public studies have less internal context. A useful benchmark would combine RIPE Atlas, browser-based measurements, open edge identifiers, and voluntary operator disclosures about regional scoping and peering changes.

A second open problem is uncertainty-aware catchment prediction. AnyOpt and related systems show that catchments can be predicted, but operators need confidence intervals, rollback criteria, and sensitivity analysis. The goal should be not only to recommend a route policy but also to state how likely it is to improve P90 latency without harming P99 latency or overloading a site.

A third open problem is IPv6 and resolver-CDN interaction. Public resolver choices, EDNS Client Subnet behavior, IPv6 preference, and CDN mapping policy interact in ways that can improve or degrade edge selection. Recent work on public DNS resolvers and CDNs shows that resolver choice can affect both DNS latency and client-to-edge mapping quality [12]. A comparative anycast study should therefore report IPv4 and IPv6 separately and identify whether DNS mapping or BGP anycast is the dominant source of latency.

A fourth open problem is operational safety under stress. Root DNS and CDNs both use anycast for resilience, but route withdrawal under attack or overload can transfer traffic to neighboring sites. The optimization problem is therefore multi-objective: reduce latency during normal operation, preserve stability during failures, and avoid changes that create hidden correlated risk.

## 10. Conclusion

Anycast performance cannot be judged outside the service that uses it. Root DNS anycast may exhibit substantial path inflation, but recursive caching and long-lived referrals usually keep root latency from dominating page-level experience. CDN anycast has less room for inefficiency because users pay edge latency repeatedly during content retrieval. This difference explains why root DNS optimization should prioritize resilience, reachability, and cache-aware latency, while CDN optimization should prioritize tail latency, peering quality, catchment control, and continuous telemetry.

The practical recommendation is straightforward: optimize anycast by objective, not by map size. Before adding sites, measure catchments. Before judging latency, weight it by application exposure. Before changing BGP policy, estimate the traffic shift and define rollback criteria. Anycast is not inherently slow or inherently optimal. It is a powerful routing abstraction whose performance depends on how carefully the operator aligns BGP behavior with application needs.